\documentclass{mn2e}
\usepackage{graphicx}
\usepackage{color}
\usepackage{url}
\usepackage[T1]{fontenc}
\usepackage{aecompl}
\footnotesize
\newdimen\minuswidth    
\setbox0=\hbox{$-$}
\minuswidth=\wd0
\catcode`@=\active
\def@{\kern\minuswidth}
\newdimen\digitwidth    
\setbox0=\hbox{\rm0}
\digitwidth=\wd0
\catcode`!=\active
\def!{\kern\digitwidth}
\normalsize
\title[ A pilot ASKAP survey of radio transient events]
{A pilot ASKAP survey of radio transient events in the region around the intermittent pulsar PSR J1107$-$5907\makeatletter
\author[G. Hobbs et al.]{G. Hobbs,$^{1}$
I. Heywood,$^{1,2}$
M. E. Bell,$^{1,3}$
M. Kerr,$^{1}$
A. Rowlinson,$^{1,3,16,17}$
S. Johnston,$^{1}$
\newauthor
R. M. Shannon,$^{1}$
M. A. Voronkov,$^{1}$
C. Ward,$^{4,1}$
J. Banyer,$^{4,3}$
P. J. Hancock,$^{5,3}$
\newauthor
Tara Murphy,$^{4,3}$
J. R. Allison,$^{1}$
S. W. Amy,$^{1}$
L. Ball,$^{1}$
K. Bannister,$^{1}$
\newauthor
D. C.-J. Bock,$^{1}$
D. Brodrick,$^{1}$
M. Brothers,$^{1}$
A. J. Brown,$^{1}$
J. D. Bunton,$^{1}$
\newauthor
J. Chapman,$^{1}$
A. P. Chippendale,$^{1}$
Y. Chung,$^{1}$
D. DeBoer,$^{6,1}$
P. Diamond,$^{7,1}$
\newauthor
P. G. Edwards,$^{1}$
R. Ekers,$^{1}$
R. H. Ferris,$^{1}$
R. Forsyth,$^{1}$
R. Gough,$^{1}$
A. Grancea,$^{1}$
\newauthor
N. Gupta,$^{8,1}$
L. Harvey-Smith,$^{1}$
S. Hay,$^{9}$
D. B. Hayman,$^{1}$
A. W. Hotan,$^{1}$
S. Hoyle,$^{1}$
\newauthor
B. Humphreys,$^{1}$
B. Indermuehle,$^{1}$
C. E. Jacka,$^{1}$
C. A. Jackson,$^{5,1}$
S. Jackson,$^{1}$
\newauthor
K. Jeganathan,$^{1}$
J Joseph,$^{9}$
R. Kendall,$^{9}$
D. Kiraly,$^{1}$
B. Koribalski,$^{1}$
M. Leach,$^{1}$
\newauthor
E. Lenc,$^{3,4,1}$
A. MacLeod,$^{1}$
S. Mader,$^{1}$
M. Marquarding,$^{1}$
J. Marvil,$^{1}$
\newauthor
N. McClure-Griffiths,$^{10,1}$
D. McConnell,$^{1}$
P. Mirtschin,$^{1}$
S. Neuhold,$^{1}$
A. Ng,$^{1}$
\newauthor
R. P. Norris,$^{1}$
J. O'Sullivan,$^{1}$
S. Pearce,$^{1}$
C. J. Phillips,$^{1}$
A. Popping,$^{11,1}$
\newauthor
R. Y. Qiao,$^{12,9}$
J. E. Reynolds,$^{1}$
P. Roberts,$^{1}$
R. J. Sault,$^{1,13}$
A. E. T. Schinckel,$^{1}$
\newauthor
P. Serra,$^{1}$
R. Shaw,$^{1}$
T. W. Shimwell,$^{14,1}$
M. Storey,$^{1}$
A. W. Sweetnam,$^{15,1}$
\newauthor
A. Tzioumis,$^{1}$
T. Westmeier,$^{11,1}$
M. Whiting,$^{1}$
C. D. Wilson,$^{1}$
\\
\\
Affiliations given at the end of the paper 
\\
\\
}}
\date{printed \today}

\begin{document}

\maketitle
\begin{abstract}
\indent 
We use observations from the Boolardy Engineering Test Array (BETA) of the Australian Square Kilometre Array Pathfinder (ASKAP) telescope to search for transient radio sources in the field around the intermittent pulsar PSR J1107$-$5907. The pulsar is thought to switch between an ``off" state in which no emission is detectable, a weak state and a strong state.  We ran three independent transient detection pipelines on two-minute snapshot images from a 13 hour BETA observation in order to 1) study the emission from the pulsar, 2) search for other transient emission from elsewhere in the image and 3) to compare the results from the different transient detection pipelines.   The pulsar was easily detected as a transient source and, over the course of the observations, it switched into the strong state three times giving a typical timescale between the strong emission states of 3.7\,hours.  After the first switch it remained in the strong state for almost 40 minutes.  The other strong states lasted less than 4 minutes.  The second state change was confirmed using observations with the Parkes radio telescope. No other transient events were found and we place constraints on the surface density of such events on these timescales.  The high sensitivity Parkes observations enabled us to detect individual bright pulses during the weak state and to study the strong state over a wide observing band. We conclude by showing that future transient surveys with ASKAP will have the potential to probe the intermittent pulsar population.
\end{abstract}

\section{Introduction}

The Australian Square Kilometre Array Pathfinder (ASKAP) telescope, currently being built in Western Australia, is designed to be a high-speed survey instrument (see Johnston et al. 2007)\nocite{johnston07}. Each antenna in the array operates with a Phased Array Feed (PAF) receiver system (Hay \& O'Sullivan 2008; Hampson et al. 2012)\nocite{hampson12}\nocite{hs08} allowing observations with a field of view of 30 square degrees.  The PAFs have a frequency range from 700 MHz to 1.8 GHz within which the correlator can process $\sim$300\,MHz of instantaneous bandwidth.  The full ASKAP telescope is not yet ready for scientific observations.  In this paper, we describe results obtained using six of the ASKAP antennas. This small array is known as the Boolardy Engineering Test Array (BETA; Hotan et al. 2014)\nocite{hotan14} and has been created to demonstrate the effectiveness of PAFs in enabling high-speed radio surveys.  BETA has a reduced capability in the number of beams that are able to be formed and therefore has a reduced instantaneous field of view (the actual field of view that was processed is described below).

One of the main goals for ASKAP is to explore the transient and variable radio sky via highly sensitive, wide-field observations. The Commensal Real-time ASKAP Fast Transients (CRAFT) survey (Macquart et al. 2010)\nocite{mbb+10} will search for fast ($<$5\,s) transient radio sources such as fast radio bursts (FRBs).  The survey for Variables and Slow Transients (VAST; Murphy et al. 2013)\nocite{mck+13} will search for more slowly varying sources ($>$5\,s) such as flare stars, intermittent pulsars and radio supernovae.  However, it is likely that unexpected transient emission will also be detected.  For instance, previous surveys (e.g., Bannister et al. 2011; Hyman et al. 2002\nocite{hyman2002}; Levinson et al. 2002\nocite{levinson2002};  Frail et al. 2012\nocite{frail2012}; Jaeger et al. 2012\nocite{jaeger2012})\nocite{bannister11} identified a number of transient and variable sources; the origin of many of these is still unknown. 

Pulsars are usually discovered through the detection of their regular sequence of pulses. However, pulsars are also known to suddenly switch off (a recent example was described in Kerr et al., 2014)\nocite{kerr14} or to emit giant pulses.  Rapidly rotating radio transient sources (RRATS; McLaughlin et al. 2006\nocite{mll+06}) are now thought to be a small number of bright pulses from otherwise undetected neutron stars.  Pulsar flux densities can also vary significantly because of interstellar scintillation or because the pulsar signal becomes eclipsed by a companion object.   The Compact Objects with ASKAP: Surveys and Timing (COAST) survey team\footnote{\url{http://pulsar.ca.astro.it/pulsar/COAST/Welcome.html}} plans to carry out traditional-style pulsar surveys using ASKAP, but the transient surveys may also find pulsars through bright individual pulses or, as described later in this paper,  by monitoring their variability in the image plane.

Our survey with BETA is not as sensitive as many previous previous transient surveys. However, we can probe areas of parameter space not covered by previous surveys  (e.g. short timescales) and we can also target fields with known bright transient sources. For example, the pulsar PSR J1107$-$5907 has an extremely variable flux density and, in its bright state, is easily detectable in our image-plane survey.  This provides us with a source that can be used to study one of the many phenomena that will arise in the main ASKAP transient surveys. It is also scientifically valuable to study this exotic pulsar in more detail. 

PSR~J1107$-$5907 is thought to have three emission states: strong, weak and off. The only in-depth study of this pulsar has been carried out by Young et al. (2014)\nocite{yws+14} who used the Parkes radio telescope to show that
\begin{itemize}
\item the pulsar remains in the strong-emission state for $\sim$200--6000 pulses with apparent nulls over time-scales of up to a few pulses at a time,
\item during the weak state a few bursts of up to a few pulses are detectable, and
\item the off-state may actually be the bottom end of the pulse-intensity distribution for the source.
\end{itemize}

Young et al. (2014) emphasised that their statistical analysis of the bright emission was limited because of their small number of detections of the strong state. We therefore conducted BETA early-science observations that were centred on PSR J1107$-$5907.   We also obtained some coincident observations of the pulsar with the Parkes 64-m radio telescope.  The goals of this work were to 
\begin{itemize}
\item identify issues relating to the search for transient radio emission in a complex region of the Galactic plane,
\item study the image quality obtained using the PAFs,
\item monitor PSR~J1107$-$5907 with much longer observation times than previous observations,
\item search for unexpected, bright transient sources elsewhere in the field,
\item compare the results from different transient detection pipelines, and
\item consider the possibility for discovering pulsars in wide-field imaging surveys.
\end{itemize}

All previous observations of this pulsar were obtained with the Parkes telescope.  In contrast, our survey (1) uses an interferometer with a wide field-of-view allowing us to study not just the pulsar, but the Galactic sources in the pulsar's vicinity, (2) is based on analysis of images instead of using traditional pulsar search methods, (3) is at a site with extremely low levels of radio-frequency interference, (4) allows us to study the pulsar with a 304\,MHz bandwidth between 700 and 1000\,MHz and (5) provides long-tracks on source ($\sim 13$ hours, compared with $\sim 3.6$ hours for the longest low-frequency observations yet carried out).  

The use of a telescope that was still being commissioned does lead to problems that are unlikely to arise in future surveys.  We note, and will discuss later, that the current array has a poor instantaneous coverage of the ($u,v$) plane.  When imaging complex regions of the Galactic plane this makes CLEANing the image challenging.  In this paper we therefore apply the transient pipelines to non-CLEANed images.  Our survey can therefore be thought of as a worst-case scenario for future transient surveys in terms of sensitivity, calibration artefacts and snap-shot image quality. 

 In \S2 we describe our observing setup with BETA and Parkes. Our results are presented in \S3.  We discuss the implications of our results for studies of the pulsar and for wide-area transient surveys in \S4. We conclude in \S5.

\section{Observations}

Our observations were centred on the position of PSR~J1107$-$5907.   The basic parameters of this pulsar, as obtained from the ATNF pulsar catalogue \cite{mhth05}, are listed in Table~\ref{fg:params}\nocite{lfl+06}. 

\begin{table}
\caption{Basic parameters from Lorimer et al. (2006) for PSR J1107$-$5907.}\label{fg:params}
\begin{tabular}{ll}
\hline
Parameter & Value  \\
\hline
Right ascension (hh:mm:ss)\dotfill & 11:07:34.46(4) \\
Declination (dd:mm:ss)\dotfill & $-$59:07:18.7(3) \\
Dispersion measure (cm$^{-3}$pc)\dotfill & 40.2(11) \\
Pulse frequency (Hz)\dotfill & 3.95611366927(8) \\
Pulse frequency time derivative (s$^{-2}$)\dotfill & $-$1.41(15)$\times 10^{-16}$ \\
Epoch of frequency determination (MJD)\dotfill & 53089 \\ 
Epoch of position determination (MJD)\dotfill & 53089 \\
\hline
{\bf Derived parameters} \\
Galactic longitude (degrees)\dotfill & 289.94 \\
Galactic latitude (degrees)\dotfill & 1.11 \\
Distance (kpc)\dotfill & 1.8 \\
Pulse period (s)\dotfill & 0.25 \\
Pulse period derivative\dotfill & $9\times10^{-18}$ \\
Characteristic age (yr)\dotfill & $4\times10^{8}$ \\
Surface magnetic field strength (G)\dotfill & $5 \times 10^{10}$ \\
Spin down energy loss rate (ergs/s)\dotfill & $2 \times 10^{31}$ \\
\hline
\end{tabular}
~\\ 
Uncertainties in parentheses are the 1$\sigma$ error on the last decimal place given for each parameter. 
\end{table}

\subsection{The BETA observations}

\begin{figure}
\includegraphics[width=8cm]{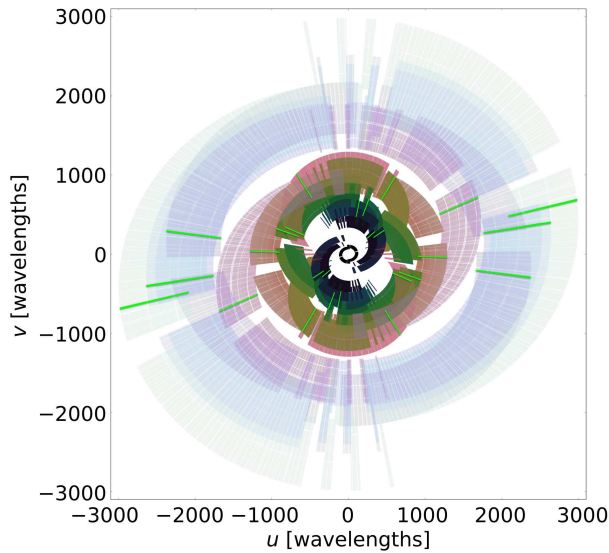}
\caption{The ($u,v$) plane coverage of the PSR~J1107$-$5907 observations. Each baseline is individually coloured.  The ($u$,$v$) plane coverage for one of the two minute snapshot images is overlaid in green. The total observing time was 13\,hours with an observing bandwidth of 304\,MHz. }\label{fg:uv}
\end{figure}

BETA has been described in detail by Hotan et al. (2014)\nocite{hotan14}. In brief, six 12-m antennas provide a total, geometric collecting area of 678\,m$^2$. The maximum baseline length is 916\,m giving an angular resolution of 1.3$^\prime$.  The beamformers were configured to deploy nine beams in a diamond pattern with a 1-2-3-2-1 row layout in declination, however for our project only the central, on-axis beam was processed. The array was configured to observe between $711.5$ and $1015.5$\,MHz with the correlator delivering 16,416$\times$18.5\,kHz frequency channels.  The integration time per visibility point is 5 seconds. 

PSR~J1107$-$5907 was observed from UTC 01:38:00 to 14:38:00 on 2014 July 14 giving a total of 13 hours of on-source time\footnote{Since the observations for this work were carried out it has been found that the time stamps were incorrect by $\sim 10$s. The times given in this paper are therefore incorrect by approximately this time (the exact offsets are not yet fully determined).  Such an offset does not change the scientific results of this paper.}. After the observation completed the array was repointed to place the strong radio source PKS~B1934$-$638 at the centre of the on-axis beam and 15 minutes of data were recorded from UTC 14:44:00 for calibration purposes. The data were subsequently averaged to form a 304$\times$1MHz channel `continuum' data set. The data are provided in standard measurement sets, and the \textsc{CASA}\footnote{\url{http://casa.nrao.edu}} package was used for editing, calibration and imaging. The \textsc{flagdata} task was used on both the target and the calibrator scans in order to remove the autocorrelations, and to clip the data (with a manually determined threshold) in order to remove occasional spurious instrumental artefacts (that give unrealistically high amplitudes or amplitudes of exactly zero). A single pass of the \textsc{rflag} algorithm with time and frequency thresholds of 5$\sigma$ was also applied to both data sets.

The PKS~B1943$-$638 scans were then averaged in time and the \textsc{bandpass} task was used to solve for the instrumental bandpass shape against the model derived by Reynolds (1994)\nocite{reynolds1994}. This set of corrections has the dual purpose of correcting the bandpass shape and setting the absolute flux density scale of the data. The derived corrections were then applied to the target data set.

At this stage the usual procedure is to apply one or more passes of self-calibration to correct the data further, with calibration of the complex antenna-beam gains being performed against a sky model derived from the image. For typical extragalactic fields away from the Galactic plane this approach works extremely well. However, self-calibration was not applied to the data presented here. With only six antennas and a maximum baseline of approximately 1\,km, BETA is a relatively sparse array. The ($u,v$) plane coverage of the target observation is shown in Figure~\ref{fg:uv}. Even with the benefits
of multifrequency synthesis imaging (e.g., Sault \& Wieringa 1994; Rau \& Cornwell 2011)\nocite{sault94}\nocite{rau2011}, significant gaps remain in the Fourier plane coverage, particularly in the inner region. Strong emission from the Galactic plane is present over a broad range of low spatial frequencies, and since the Fourier domain is somewhat sparsely sampled by the interferometer the large scale emission is not faithfully reproduced in the image domain. This precludes self-calibration: using the image to derive a sky model results in one that is highly incomplete and does not accurately transform into an approximately true representation of the visibilities. We trialled different methods in which we used subsets of the available baselines and only using compact features, but these methods made the image noticeably worse, suppressing many of the real diffuse structures. Thus the images that are presented in this paper have been produced with the complex gain corrections derived from the single observation of PKS B1934$-$638 as their sole correction.   However, as will be shown below and in the next sections, the images produced accurately reproduce features known in the field and allow us to successfully detect the pulsar.

\begin{figure*}
\includegraphics[width=17cm]{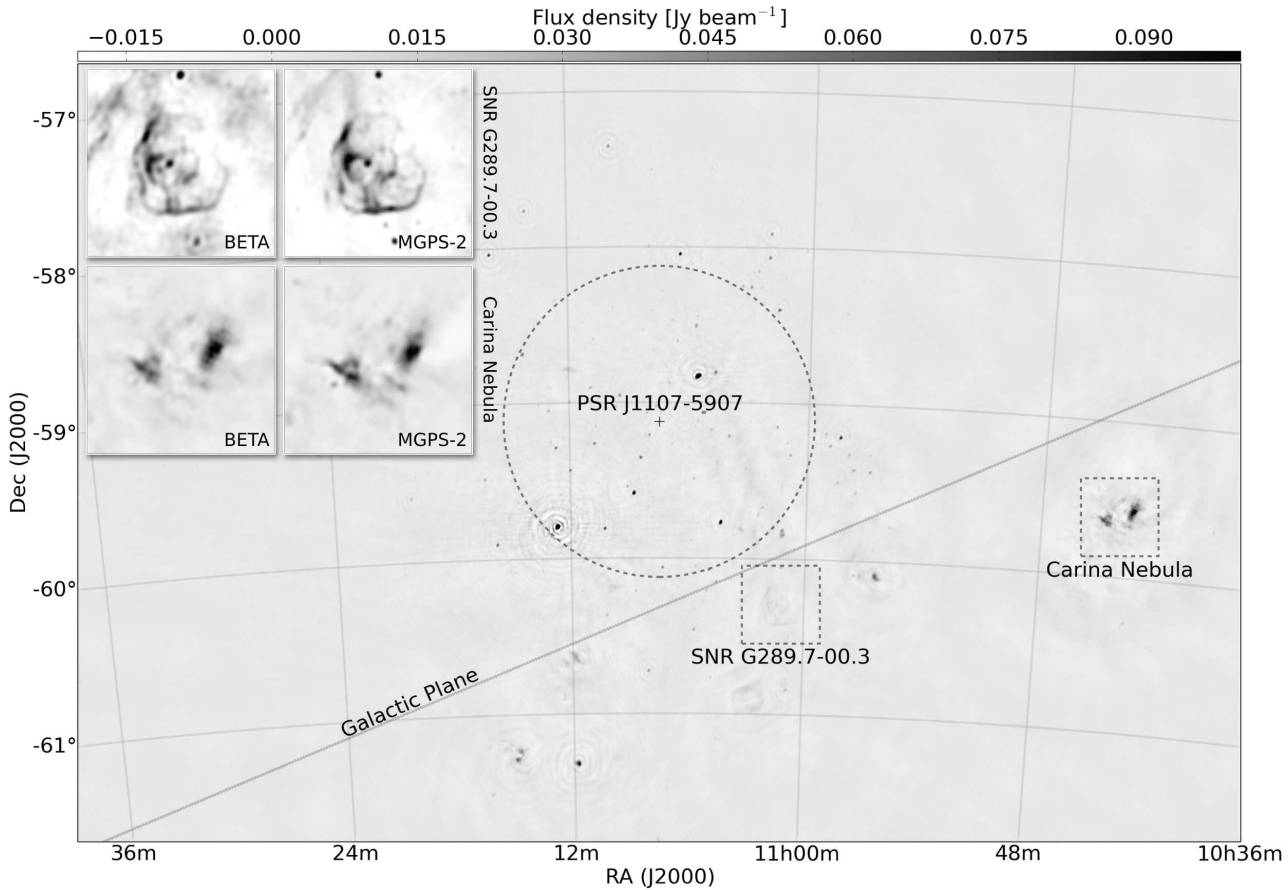}
\caption{BETA image of the area around PSR~J1107$-$5907 formed from the central on-axis beam with a central frequency of 863.5\,MHz. No primary beam correction has been applied. The array pointing position is that of the pulsar, which is marked with the cross. The dashed circle shows the approximate half power point of the formed PAF beam at the centre frequency.  Inset are two examples of complex structures in the Galactic plane that BETA images (left) with excellent correspondence to the existing Molonglo Galactic Plane Survey (MGPS-2) images. (right) The intensity scale of the main image is given by the grey-scale bar at the top of the figure. Note that the inset images are not shown on the same brightness scale due to the primary beam attenuation. The Carina nebula is observed through the first sidelobe of the primary beam. }\label{fg:finalImage}
\end{figure*}

The data from the beam centred on PSR~J1107$-$5907 were imaged using the \textsc{CASA clean} task. A 2240$\times$2240 pixel image was produced with each square pixel spanning 12 arcseconds, giving a total sky area of 7.5$\times$7.5\,degrees. The image was
deconvolved interactively employing a multifrequency synthesis algorithm that models each component with a 3$^{\rm rd}$ order Taylor series (Rau \& Cornwell 2011), manually improving the cleaning mask in order to avoid spurious features with each major cycle. The final deconvolved wide field image is presented in Figure~\ref{fg:finalImage} and described in Section~\ref{sec:results}. No primary beam correction has been applied.

Searches for transient radio emission can be based on the raw visibilities obtained from the telescope or on snap-shot images.  It is likely that transient searches (for events lasting more than a few seconds) with the full ASKAP array will be based on studies of snap-shot images and so we use the same procedure here. 
Before these images were produced, the spectral clean component model produced as a result of the imaging step was inverted into a set of model visibilities. This model was then subtracted from the observed data to produce a residual visibility database (note that the pulsar was not included in this model). The residual visibilities were then split into two-minute intervals (390 epochs) and dirty (i.e., not cleaned) images were formed to be processed by the transient search pipelines.   The choice of two-minute intervals was based on the need to study the shortest time scale variability possible, whilst ensuring that the pulsar can be detected with high S/N in its bright state.

In summary, our transient search is based on two-minute snap-shot images. In those images the brightest sources have been removed as well as possible using the CLEAN component model that was derived from the full-length image.  The CLEAN procedure was not applied to the snap-shot images. Any transient source that lasted up to a few minutes would therefore be detectable in the snapshot images with the shape of the dirty beam at that time. The images described in this paper are available for download from the CSIRO data archive\footnote{\url{http://hdl.handle.net/102.100.100/25153?index=1}.  Relevant files have filenames beginning with SB220 (this nomenclature is from the scheduling block \#220).}.

\subsection{The Parkes observations}

The main aim of our work is to use PSR~J1107$-$5907 to demonstrate the effectiveness of transient searches with ASKAP. In order to confirm our results, and to obtain multi-frequency observations of the pulsar, we carried out simultaneous observations with the Parkes telescope.  The Parkes telescope has a long history of detecting and studying pulsars and fast-transient events and relatively straight-forward and mature procedures are in place to record and process the observations. The Parkes observations described here were carried out as part of a program to study the stability of the pulsed emission (project code P863\footnote{After the standard embargo period of 18 months these data will be publically available for download through the CSIRO Data Access Portal; \url{http://data.csiro.au} (Hobbs et al. 2011)\nocite{hobbs2011}. The relevant files are s140714\_095334 and t140714\_09333.}) through long, multi-frequency observations of a small sample of pulsars (which includes PSR~J1107$-$5907). These observations were obtained in ``search-mode", in which the data are averaged over a short integration time (256$\mu$s) from UTC 09:53 to 10:53 on 2014 July 14.  We observed using a dual-band receiver providing 64\,MHz of bandwidth centred on 732\,MHz and simultaneously 1\,GHz of bandwidth centred on 3100\,MHz.

Before the pulsar observation commenced we observed a pulsed-calibrator signal for subsequent polarisation and flux-density calibration. The band edges have low-gain and are often corrupted and so, following standard procedure, we removed the outermost 5\% of the channels. A recent flux calibration file for the appropriate receiver and backend was selected from archival data. A \textsc{psrsh} script with instructions to load these files and to undertake frontend and flux calibration was produced. This script was loaded into the \textsc{dspsr} software package along with an ephemeris for the pulsar to output single calibrated pulse profiles with 256 phase bins and 512 frequency channels. The \textsc{dspsr} package was similarly used to produce time averaged profiles with two minute integrations. The two minute integrated profiles were frequency-averaged. The \textsc{psrflux} package with self-standardisation was then used to estimate the average flux density of each two minute profile and its uncertainty.  We have also studied the entire time series (i.e., without splitting into individual pulses) using the \textsc{pfits} software package.  For this processing we summed the two polarisation states, dedispersed at the known dispersion measure of the pulsar and then summed in frequency.  We then do not flux calibrate these time series.

\section{Results}\label{sec:results}

\begin{figure*}
\includegraphics[width=14cm]{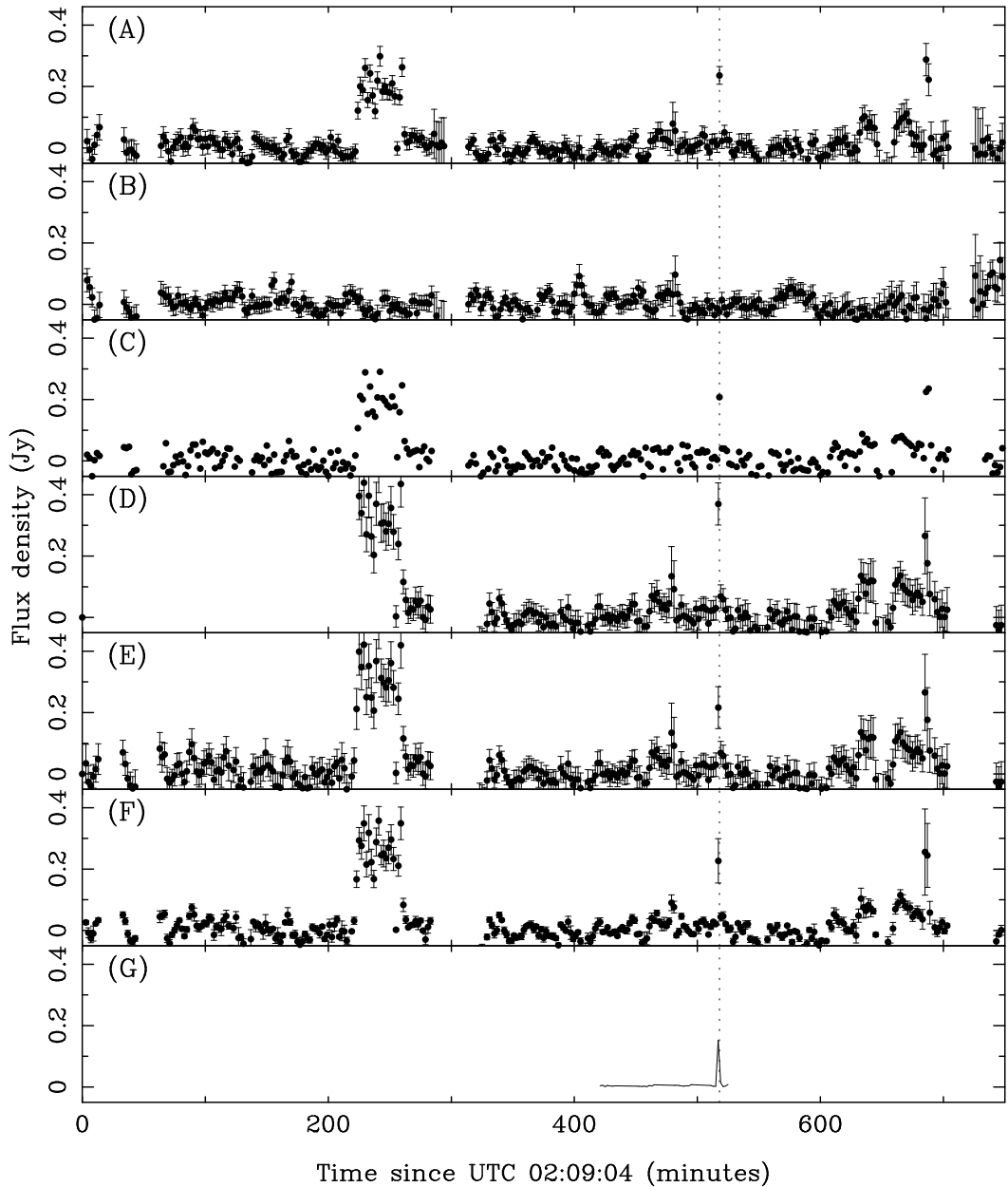}
\caption{(A) Flux density versus time for PSR J1107$-$5907 and (B) for a sky position close to the pulsar as obtained directly from the original images. (C) shows the results from Pipeline~1. Panels (D) and (E) show the  blind-search and monitored-search respectively output from the TraP pipeline.  Note that panel (D) shows the direct output from the TraP pipeline, which only provides data after the transient is detected.  Panel (F) contains the output from the VAST pipeline.   (G) shows the flux density obtained in the 40\,cm band from the Parkes telescope.}\label{fg:day1}
\end{figure*}

As this paper reports some of the first observations carried out by ASKAP, we first confirmed that the image contains known sources and that the flux and astrometric calibration was adequate for our purposes. An overview of the area around PSR~J1107$-$5907 is shown in Figure~\ref{fg:finalImage}. The position of the pulsar is marked, and the dashed circle represents the approximate half power point of the formed beam at the centre of the band. The Galactic plane runs diagonally across the image. The pixel scale is given by the grey-scale bar at the top of the image and saturates black at 100\,mJy\,beam$^{-1}$.

The broad ripples introduced by the undersampled Galactic emission are evident, however smaller scale individual features in the plane of the Galaxy are reproduced remarkably well. Inset into the upper left of Figure~\ref{fg:finalImage} are two such examples, featuring the BETA image (left panels) and comparison images from the second epoch Molonglo Galactic Plane Survey (Murphy et al. 2007, MGPS-2; right panels)\nocite{murphy2007}. The upper row shows the supernova remnant SNR~G289.7$-$00.3 (e.g., Green 2009)\nocite{green2009} and the lower row shows the bright Carina nebula (e.g., Preibisch et al. 2011)\nocite{preibisch2011}. There is excellent correspondence between the features recovered by ASKAP and those seen in the MGPS-2 images. Note however that these images are not on the same intensity scale. For SNR G289.7$-$00.3 the ASKAP and MGPS-2 images saturate at 13 and 50\,mJy\,beam$^{-1}$ respectively, and the corresponding values for the Carina nebula are 0.1 and 1.6\,Jy\,beam$^{-1}$. We note that the Carina nebula has been imaged through the first sidelobe of the primary beam\footnote{The ASKAP antennas can rotate to keep the parallactic angle fixed as an observation progresses, thus the sidelobes do not rotate with respect to the sky, and a typical ASKAP observation generally detects a significant number of sources in the sidelobes without distortion.}.

As the position of the pulsar is known, we simply measured the flux density corresponding to the known source position as a function of time using the 2\,minute snap-shot images.  This is shown in panel (A) of Figure~\ref{fg:day1}.  In panel (B) we show the flux density at a nearby position in the sky.  The pulsar was undetectable for most of the observation. However, three events occurred (around 200, 500 and 700 minutes from the start of the observation) in which the pulsar switched into its bright state and flux densities of around 200\,mJy were observed.  Panel (B) shows that remaining imaging artefacts do vary the baseline noise level, but the pulsar state changes are easily detectable.

\subsection{Transient detection pipelines}

Even though the pulsar state changes were detected using the simple method described above, we also applied various transient pipelines to the data set.  The aim was to 1) study the effectiveness of those pipelines with this challenging data set and 2) to search for unexpected events elsewhere in the image.  
We trialled three pipelines. The first pipeline was developed specifically for this data set.  The second pipeline was developed for the LOFAR telescope (the Transient Pipeline; TraP; Swinbank et al. 2015)\nocite{swinbank2015}.  The third pipeline was developed to search for transient events with the completed ASKAP telescope (the VAST Transients Pipeline; Murphy et al. 2013).

Both the TraP and VAST pipelines have been designed to work with CLEANed images. Sources that have not been de-convolved are therefore problematic for the pipelines. However, because a sky model has been subtracted from these snapshots, any point-like flux found in these images is either a transient object or an imaging artefact. In either case, the pipelines offer convenient ways to identify and explore these objects.

\subsubsection{Pipeline 1}

The first pipeline uses a straightforward matched-filter methodology in which the point-spread-function of the beam is matched with every pixel of the image.  To mitigate edge effects, we considered only the central quadrant (1024 pixels).  At each position within this quadrant, we computed the log likelihood difference, $\delta\log{\mathcal{L}}$, between the null and alternative hypotheses representing the absence and presence, respectively, of a source.  If the image noise were Gaussian, the significance of excesses would be given by $\sqrt{2\delta\log\mathcal{L}}$.  However, the noise in the image is dominated by sidelobes of the point-spread-function, thus is highly correlated, non-Gaussian, and time-dependent.  We thus adopted a ``normalised'' statistic, $S = \delta\log{\mathcal{L}}/\sigma^2$ where $\sigma^2$ is the root-mean-square deviation for each residual image.

\begin{figure}
\includegraphics[width=6cm,angle=-90]{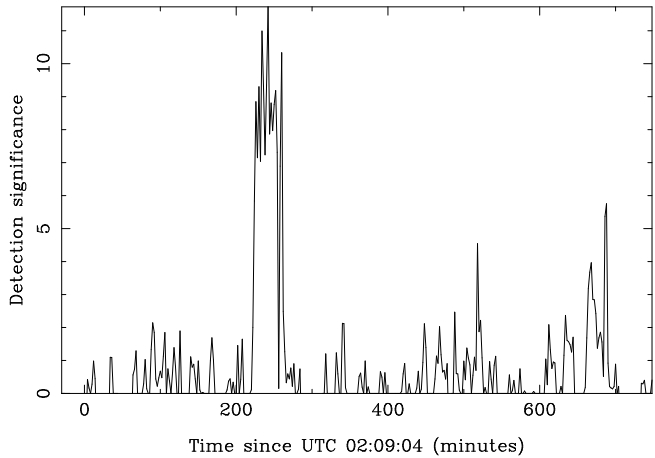}
\caption{Detection significance as a function of time at the position of the pulsar from Pipeline 1.}\label{fg:kerrSignificance}
\end{figure}

To determine a threshold for detection, we `self calibrated' by examining the distribution of $S$ in an 80 x 70 pixel subset of the images.  Interestingly, we found that, when normalized to the mean value of $S$ within this region, the overall distribution of $S$ follows the theoretical $\chi^2$ distribution well to significances $>$5$\sigma$.   In panel (C) of Figure~\ref{fg:day1} we plot the flux density of the pulsar as measured using this pipeline as a function of time. The three burst events are clearly detectable.  This pipeline does not naturally provide an error bar for each flux density measurement.  Instead we obtained the significance of the event being real. These significance values are shown in Figure~\ref{fg:kerrSignificance}.  The first bright state is detected with high significance ($\sim10\sigma$) in each two minute bin.  The second and third bursts are detected with significance values of $\sim5\sigma$.  

\begin{figure}
\includegraphics[width=10cm]{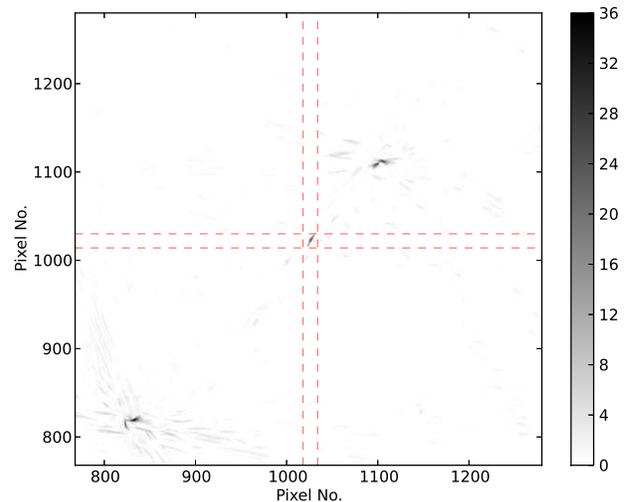}
\caption{The grey-scale bar indicates the number of times a given pixel in the image has been significantly detected by Pipeline 1. The pulsar is at the centre of the image.  The sources to the top-right and to the bottom-left are related to the incompletely subtracted bright sources (MGPS J110538$-$584950 and MGPS J111245$-$594729).}\label{fg:threshold}
\end{figure}

  In order to search for other burst events in the field we choose $S>25$, nominally $5\sigma$, as the threshold for source detection.  Using the `known' flux of the pulsar, this threshold translates to a flux sensitivity of about 65 mJy for a two-minute snapshot.  This sensitivity rises slightly as the ($u,v$) coverage becomes poorer at low source elevations. We summed the $S$ images obtained from each snapshot to produce a significance map of the field.   We next counted the number of pixels surpassing the $5\sigma$ threshold and show the results in Figure~\ref{fg:threshold}.  The pulsar (which is significant in $\sim$20 snapshots) and the artefacts produced by the incompletely subtracted bright sources dominate the field.  We therefore conclude that the pulsar can easily be detected using this algorithm and that no other unexpected transient sources exist in the data.
  
 The application of a matched filter, pixel-based technique to the dirty images is not common in radio transient searches, but was well-suited to the quality of these commissioning data.  We expect image quality with the full ASKAP array to be substantially better, both due to improved UV coverage and improved calibration and beamforming techniques.  Searches for transients in such data will certainly benefit from cleaned images, particularly those in regions requiring complicated sky models.  On the other hand, fully cleaning images is substantially more expensive from a computational standpoint.  For transient searches of data based on short (minute time-scale) integrations, a pixel-based method like the one used here could provide an efficient, quasi-real time transient search capability that is optimally matched to instantaneous UV coverage of the array.

\subsubsection{The LOFAR Transients Pipeline}

The LOFAR Transients Pipeline (TraP) has been developed by the LOFAR Transients Key Science Project in order to search automatically for transient and variable sources (Swinbank et al., 2015)\footnote{The TraP software~\cite{trap2014} is publically available: \url{https://github.com/transientskp/tkp} with documentation at \url{http://docs.transientskp.org}.}. We applied release 2.0 of this pipeline to our data set.  Initially the pipeline determines the RMS noise in the inner quarter of each residual image.  The noise was found to be roughly Gaussian with an RMS noise of 21\,mJy for the 2\,minute residual images.

We assumed that any transient event will occur from a point source and, in these images, a point source will take the shape of the point spread function (PSF).   We performed an unconstrained source extraction on the PSF of the dirty beam and use the result from a Gaussian fit to the most significant part of the PSF as the ``restoring beam shape". Sources were blindly extracted from the images using a 5$\sigma$ detection threshold and a 3$\sigma$ analysis threshold\footnote{The detection threshold is the threshold above which sources are considered ``found".  The pipeline then extracted all the pixels around a detection to the 3$\sigma$ level.  A Gaussian function was fitted to those 3$\sigma$ and above pixels to determine the source flux density. Full details are given in Swinbank et al. (2015).}, fitted to the shape of the most significant part of the PSF. The position of the sources can vary significantly in these images because of systematic calibration errors.  Therefore, we used a 5\,arcmin systematic position uncertainty to aid with source association.  All other TraP parameters were kept as the default values and we searched a FoV of 6.1$\times$6.1\,degrees centred on the phase centre of the image. 

Newly detected sources were identified within TraP.  A source is considered to be a transient if the measured flux density was greater than 8 times the worst RMS noise region from the previous best image.  This $8\sigma$ threshold was determined from the $5\sigma$ detection threshold and a 3$\sigma$ margin. 

As our images are ``dirty images", TraP finds an excess of imaging artefacts extracted around the location of the two, incompletely subtracted, brightest sources.  With the exception of PSR~J1107$-$5907, which is blindly extracted in multiple images, no convincing transient or variable sources were identified. Four sources were labeled as confirmed transients by the pipeline: PSR~J1107$-$5907, an artefact from a bright sidelobe of PSR~J1107$-$5907 and two imaging artefacts from one of the subtracted sources. 
The results from the blind search are shown in panel (D) of Figure~\ref{fg:day1}.  Additionally, TraP has a monitoring functionality where the fit is always conducted at the position of interest.  We re-ran the pipeline utilising the average position from TraP of PSR~J1107$-$5907. The resulting light curves obtained using the two minute snap-shot images are shown in panel (E) of Figure~\ref{fg:day1}. This pipeline clearly identifies the initial large burst occurring around UTC 05:53:04 which lasts for $\sim$40\,minutes.  As expected, the pipeline also detects statistically significant bursts at UTC 10:47:04 and 13:35:04.   These two panels (D and E) are almost identical.  The exception being the second ``on" period around 500 minutes after the observation start.  The flux density measured in panel (D; the blind-search method) is significantly higher than that in panel (E).  This is because the apparent position of PSR~J1107$-$5907 slightly varies throughout the observation (with a maximum deviation of $\sim$13\,arc\,seconds). This variation is caused by a combination of instrumental and calibration effects, the low S/N determination of the actual source position in the 2-minute snapshot images and ionospheric variations.   In panel (D) a fit has been made for the flux density of the closest source that can be associated with the pulsar.  This is therefore a better measure of the pulsar's flux density.

\subsubsection{The VAST pipeline}

The Variables and Slow Transients (VAST) detection pipeline is fully described by Bell et al. (2014)\nocite{bell2014}. In brief, for a time-sequence of images the VAST pipeline firstly performs a source finding procedure via the AEGEAN algorithm (see Hancock et al. 2012\nocite{hancock2012} for details). We use a background estimation tool to efficiently determine local measurements of the noise in regions surrounding the sources. 
Sources are then cross-matched as a function of time (and frequency) and variability statistics are calculated for the resulting light curves. 

This pipeline was applied to our data set and we followed the same procedure of calculating the PSF as described in the previous subsection. The pipeline identifies the pulsar as a transient object and the resulting light curve is shown in panel (F) of Figure~\ref{fg:day1}. This pipeline clearly shows that all three burst events from the pulsar are detected. No other sources in the field were identified as being transient. A number of artefacts were identified within the images owing to poorly subtracted bright sources, lack of ($u,v$) coverage and deconvolution: these artefacts were easily mitigated by visual inspection.

\subsection{The Parkes observations}

Panel (G) in Figure~\ref{fg:day1} shows the flux density measured at 40\,cm (732\,MHz) from the Parkes telescope for observations that have been averaged in 2\,minute intervals.  The total observation time is much shorter than that obtained using ASKAP.  However, the Parkes data clearly show that the pulsar switched to its bright state at a time that agrees with the ASKAP observations and confirms that the second ``on" state is real. The flux density during the ``on" state when averaged over 2\,minutes is 150\,mJy.  This is slightly lower than that obtained with ASKAP, but consistent with the uncertainties on the ASKAP flux density measurements. We note that the Parkes data is at a slightly lower frequency than the ASKAP observations and therefore it is not surprising that the flux density measurements do not perfectly agree.

\section{Discussion}

\subsection{The pulsar}

The observations described here were undertaken to study the effectiveness of transient pipelines for identifying strong radio burst events. The ASKAP observations provide the longest observations to date of the pulsar.  We have 11.2\,hours of data in which we can search for transient emission (for this calculation we do not account for the the data near the start and near the end of the observation in which some data are missing). During this time the pulsar was in the strong state for $\sim44$\,min giving an ``on"-fraction of $\sim6.5$\%.  This agrees well with the result of Young et al. (2014) of 6\% obtained with Parkes data alone.   These switches give a typical time scale for the bursting phenomenon of $11.2/3 = 3.7$\,hours although we emphasise that the event times are not periodic.

Young et al. (2014) noted that the strong-mode pulses are only emitted during relatively short burst periods (they report intervals of $\sim60$\,s to $\sim24$\,minutes).  The ASKAP observations show that the interval can be even longer.  The initial strong-mode lasted around 44 minutes.  However, there is at least one two minute sub-image during that time in which the measured flux density is close to the baseline level suggesting that the pulsar switched off for up to 2 minutes before returning to the bright state.

\begin{figure*}
\includegraphics[width=12cm,angle=-90]{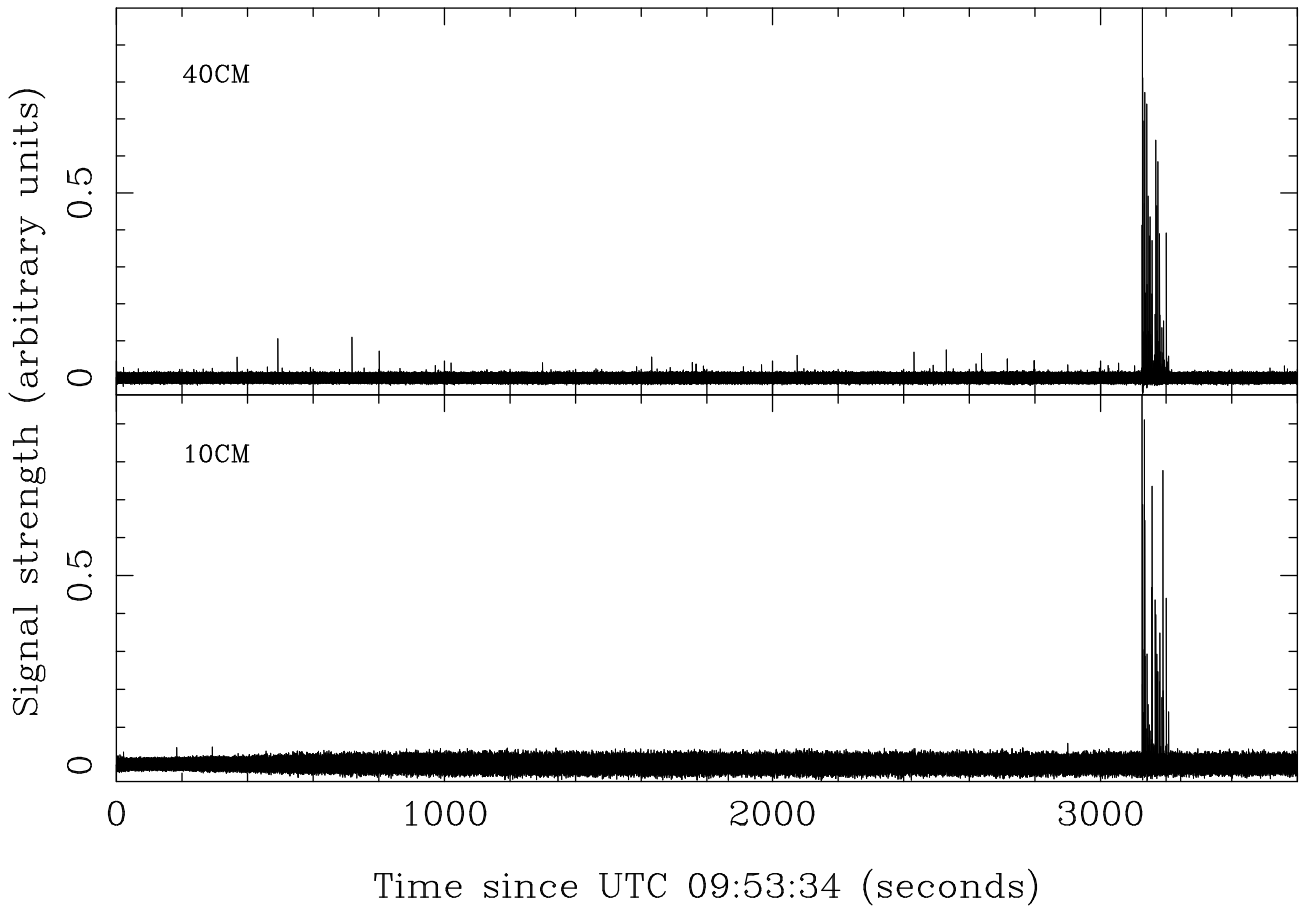}
\caption{The Parkes observation of the single pulses from PSR J1107$-$5907.  The top and bottom panels show the 40\,cm and 10\,cm observations respectively}\label{fg:fullSeries}
\end{figure*}

\begin{figure*}
\includegraphics[width=12cm,angle=-90]{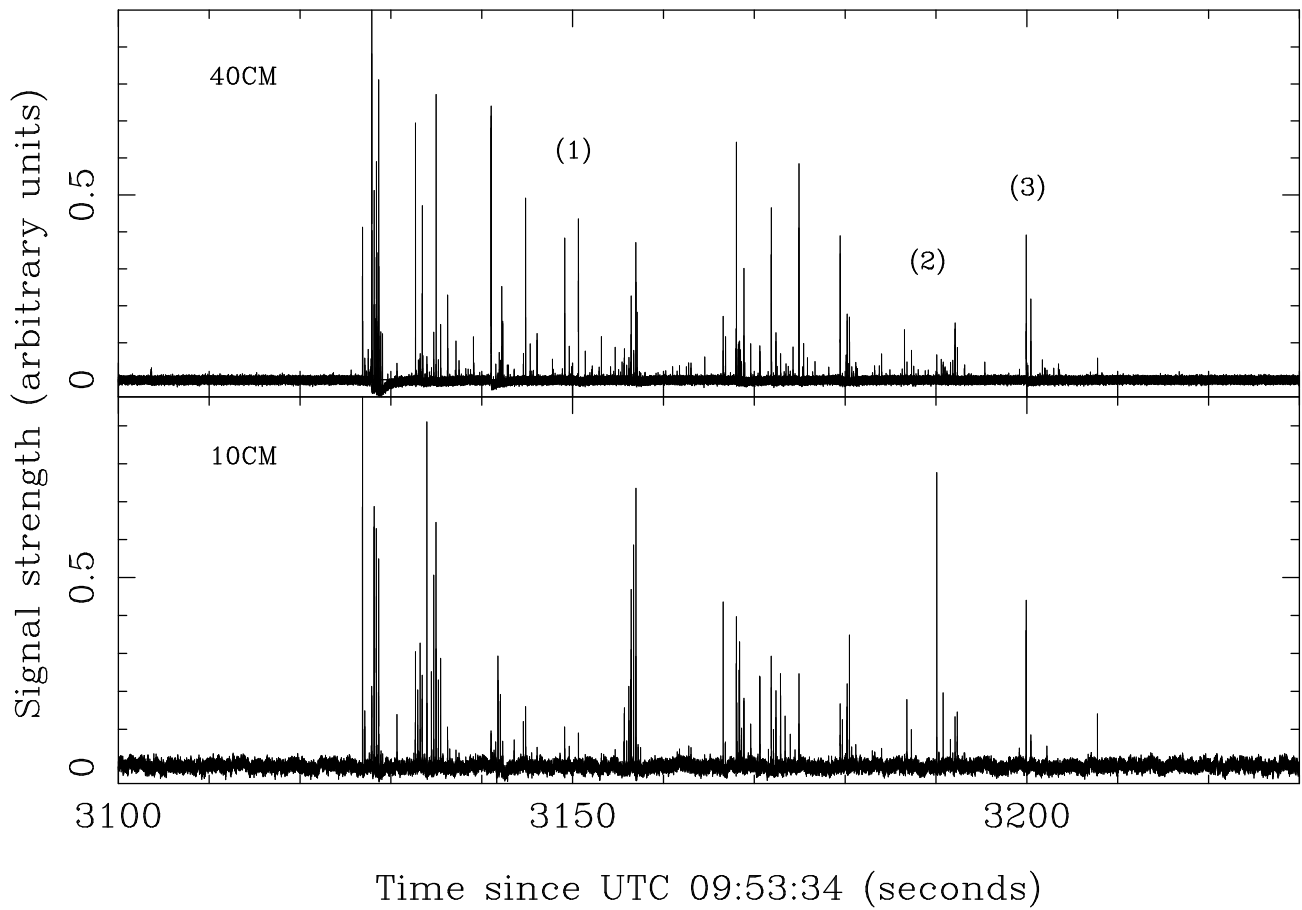}
\caption{The Parkes observation of the single pulses from PSR J1107$-$5907 during the time that the pulsar remained in the strong emission state.  The top and bottom panels show the 40\,cm and 10\,cm observations respectively}\label{fg:zoom}
\end{figure*}

The high sensitivity of the Parkes telescope and the use of a dual-band receiver system provides the opportunity to study the pulsar in more detail than before.  The Parkes sensitivity is such that we can detect individual pulses from the pulsar in its bright state and, via averaging over a few minutes, can detect the pulsar in its weak state.  For instance, we show in Figure~\ref{fg:fullSeries} the time series recorded with Parkes in both observing bands.  We have not flux calibrated these data streams, but it is clear that the pulsar switched into the bright state near the end of the Parkes observation. 

In Figure~\ref{fg:zoom} we re-plot the Parkes observations around the region of the bright burst event.  Individual pulses from pulsars are known to be complex and exhibit significant pulse-to-pulse variability.   This is clearly true for PSR~J1107$-$5907, but with the high S/N Parkes observations of the single pulses we can study the pulse-to-pulse variability across a wide-bandwidth.   We have highlighted three pulse regions in the figure.  Pulse region \#1 is significantly brighter in the 40\,cm band than in the 10\,cm band.  This is typical of most pulsar pulses - pulsars typically have steep spectral indices and are brighter at lower frequencies.  However, in contrast pulse \#2 is significantly brighter in the 10\,cm band and almost undetectable in the 40\,cm band.  Pulse \#3 has a similar flux density in the two bands.  These data sets have not been flux calibrated and so we cannot determine an absolute spectral index, but it is clear that the spectral index is significantly changing between the individual pulses. 

 We therefore conclude:

\begin{itemize}
\item The pulsar switched into the bright state at exactly the same time in the two observing bands.
\item The pulsar was in the bright state for around 81 seconds (322 pulses), but the pulses are not constant during the bright state.  Many pulses seem to be missing and the flux density of the observed pulses vary dramatically.  We have inspected the dynamic spectra for some of the brightest single pulses.  These dynamic spectra show that the diffractive bandwidth for J1107$-$5907 is  smaller than the observing bandwidth in these observing bands and so the flux density variations are not caused by the interstellar medium.
\item The spectral index for individual pulses varies significantly.  
\end{itemize}

\begin{figure*}
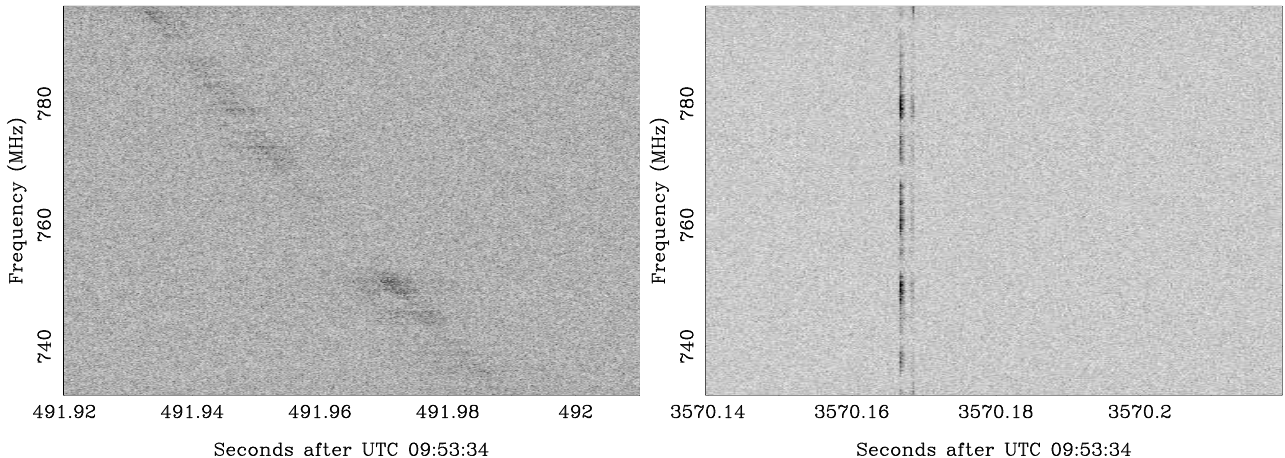

\includegraphics[width=6cm,angle=-90]{pulse_example.ps}
\includegraphics[width=6cm,angle=-90]{rfi_example.ps}
\caption{Example pulses occurring when the pulsar is in the weak state. The left-hand pulse originated at the pulsar. The right-hand pulse was caused by radio-frequency-interference (RFI).}\label{fg:examplePulses}
\end{figure*}

The 40\,cm data set in Figure~\ref{fg:fullSeries} contains a few individual bright pulses that occur outside of the strong emission state. However, this observing band is affected by radio-frequency-interference (RFI). The RFI can be easily distinguished from pulses originating at the pulsar by identifying dispersed pulses.  In Figure~\ref{fg:examplePulses} we show two examples.  The first is clearly a dispersed pulse originating from the pulsar and the second is RFI.  In the 40\,cm observing band, the dedispersed time series indicated 67 significant pulses in the ``weak" or ``off" states.  We inspected each of these by eye and found that 14 were clearly RFI, 13 were definitely pulses from the pulsar and the remainder were too weak to make a definitive statement.  We note that no clear bright pulses from the pulsar were detected after the end of the strong state (for the final 380\,s of our Parkes observation).  This may suggest that the pulsar became ``quieter" after the strong emission state ceased, or this this may simply be a co-incidence (there are earlier times in which the pulsar exhibited no detected pulses over a similar duration).

In this section we have provided an initial look at the high-time-resolution data being obtained by the Parkes P863 observing program. This program is continuing and a detailed study of the individual pulses from this pulsar using Parkes observations that span many years will be published elsewhere.

\subsection{Comparison of the pipelines}

Pipeline 1 was designed for this data set. The other two pipelines were designed for much cleaner data sets.  The first pipeline searches for pixels that change between different images. The TraP and VAST pipelines are based around identifying and tracking sources in the image.  If the flux density of any source significantly changes then that source is identified as a possible transient. Whereas pipeline 1 was not optimised for speed, the other pipelines are required to run on very large data sets and therefore are required to run quickly on a given data set.

Within the TraP and VAST pipelines there are subtle differences in the underlying algorithms and techniques used to produce results. Both pipelines, however, have a workflow that is fundamentally similar and the final science goals are the same. So it is unsurprising that both of these pipelines were able to clearly detect PSR~J1107$-$5907 and that the light curves in Figure~\ref{fg:day1} (D), (E) and (F) are similar. 

There are differences in how the VAST pipeline and TraP deal with a newly detected transient source such as PSR~J1107$-$5907. After detection of a new source within an image, both pipelines operate in the following manner for subsequent images (in time). If a given source is blindly detected in multiple images it will be cross-matched but, if it is undetected, then a flux measurement will be taken at the location of the source using the restoring beam parameters (or, as used for this paper, the PSF shape). The difference between the pipelines lies in the treatment of images prior to the transient event. Upon the first detection of a new source, the VAST pipeline will go back through all the previous images and measure the flux at the location, thus building up a complete light curve of the source and confirming its transient nature. However, refitting the source in all the previous images can be very time consuming when there are hundreds of images. Additionally, in some cases the previous images may have been discarded (because of data storage restrictions), so they will not be available for analysis. TraP has been designed to counter these two issues by assuming that the images are not available and uses statistics from the previous images, stored in the image database, to determine if the new source is a transient (as explained in Swinbank et al. 2015). Therefore, TraP outputs the light curve of transient sources from the time of first detection (as shown in panel D of Figure 3). In addition to the blind searches, TraP has the functionality of a monitoring list, in which the flux is always measured at the position of specified sources irrespective of the blind detections. As these observations were specifically targeting the variability of PSR~J1107$-$5907, we were able to take advantage of this monitoring list functionality to extract the full light curve of PSR~J1107$-$5907 (as shown in panel E of Figure 3), which can be directly compared to the outputs of Pipeline 1 and the VAST pipeline.

We note that the flux density uncertainties obtained by the VAST pipeline and TraP are different, with larger uncertainties quoted by TraP. This is likely related to the differences in fitting approaches and background characterisation used by the source finders within each pipeline and the constraints enforced on each of the fits (position and Gaussian shape). For instance, TraP always assumed a point source and all blind fits by the source finder were constrained to Gaussians taking the shape of the PSF, while the VAST pipeline used fully unconstrained fits on blind detections. Both source finders assumed the identical constraints for fits when the source was undetected. As neither source finder has been designed to deal with unCLEANed images of this type, a detailed comparison would be best conducted with an appropriately CLEANed dataset and this is beyond the scope of this paper.

\subsection{Implications for transient sources}

Even though BETA has relatively poor sensitivity compared with other transient surveys (e.g. Carilli et al. 2003\nocite{carilli2003} and Mooley et al. 2013\nocite{mooley2013}), we have studied a relatively large sky area on a time-scale that has not been probed in great detail before in the 20\,cm observing band (from 2 minutes to 12 hours).  For instance, large single dish telescopes such as the Parkes telescope have probed millisecond to second time scales over periods of a few hours, often for a single object. The majority of transient imaging surveys focus on studying timescales greater than minutes, typically days to months (see Table~5 in Carbone et al. 2014\nocite{carbone2014} for details). These imaging surveys, so far, have typically operated between $1.4 - 5$~GHz and quite rarely achieve sensitivities below 1~mJy. Note, a number of blind transient surveys at frequencies $<$500~MHz have opened up the low end of the radio spectrum to transient detection (see Lazio et al. 2010; Jaeger et al. 2012; Bell et al. 2014; Carbone et al. 2014)\nocite{lazio2010}\nocite{jaeger2012}; ASKAP will be able to detect transients at intermediate frequencies from 700 to 1800 MHz. 

\begin{figure}
\includegraphics[width=9cm]{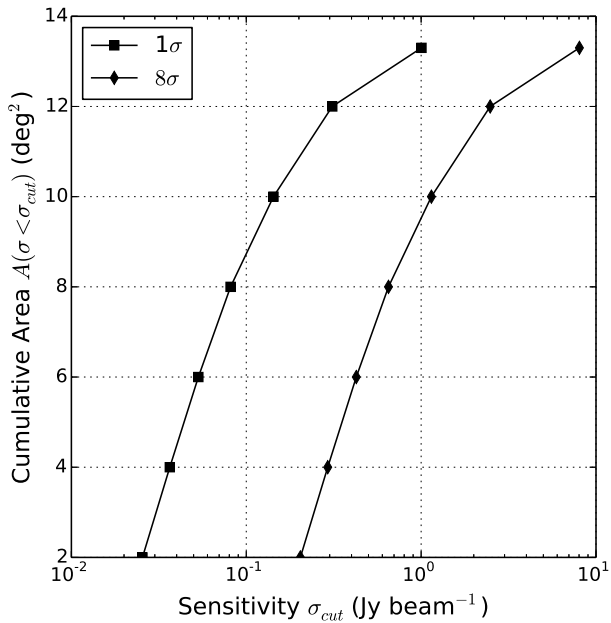}
\caption{Log sensitivity $\sigma_{cut}$ ($\mu$Jy beam$^{-1}$) versus cumulative area $A$ (deg$^{2}$) normalised over all epochs. The cumulative area gives the amount of sky we have observed with sensitivity $\sigma < \sigma_{cut}$. For example, approximately 2.0 deg$^{2}$ was observed with sensitivity better than 25~mJy.} 
\label{fg:thresholdDet}
\end{figure}

In this section we consider the rate of transients events as none were detected. We exclude the pulsar from this analysis as it was a targeted source. Presenting and comparing results from transient surveys is non-trivial as different surveys probe different timescales, observing frequencies and sky areas. To ascertain the area and sensitivity this survey has probed we perform the following analysis (note this differs from the detection threshold reported for pipeline 1). 
\begin{itemize}
\item For each image we calculate an RMS map where each of the pixels represents the 1$\sigma$ sensitivity in that image and region.
\item A Gaussian primary beam correction is applied to the RMS map values. Note, the ASKAP-BETA primary beam response has not been fully characterised yet. We therefore work under the assumption that each antenna is a fully illuminated circular aperture with Gaussian profile.  
\item  Over all the images we calculate the cumulative area that the images probe for a number of different sensitivity bins. We normalise this cumulative sky area by the number of images in our sample and we plot the results in Figure~\ref{fg:thresholdDet}. We plot both the 1$\sigma$  values and also the 8$\sigma$ values of sensitivity. The $8 \sigma$ curve represents the detection threshold of this survey. We will discuss further below caveats to this analysis. 
\end{itemize}

We can use Poisson statistics to calculate an upper limit on the expected number of transient events (per deg$^{2}$) in any given snapshot at this frequency, sensitivity and cadence (see also Bell et al. 2014). This snapshot surface density ($\rho$) is defined as:

\begin{equation}
\rho < \frac{\ln(0.05)}{(n-1) \times \Omega}, 
\label{upp_eq}
\end{equation}
  
\noindent where $\Omega$ is the total field of view (or sky area) and $n$ is the number of independent images searched ($n=389$).  

From Equation \ref{upp_eq} we therefore obtain the upper limit on the snapshot surface density of transient events calculated as function of area and sensitivity, listed in Table \ref{area_sens_table}. At the most sensitive part of the images ($<2$\,deg$^{2}$) we place a limit of $\rho < 3.9 \times 10^{-3}$\,deg$^{-2}$ above a detection threshold of 0.2\,Jy~beam$^{-1}$. Over the full field of view that was searched (13.3\,deg$^2$) a limit of $\rho < 5.8 \times 10^{-4}$\,deg$^{-2}$ above a detection threshold of 8\,Jy~beam$^{-1}$.
 
The half-power beam width is defined as $\theta = 1.028 (\lambda / D)$ and assuming $\lambda=0.34$m (centre frequency of the observations) and $D = 12$m we find an area of A = $\pi(\theta /2)^{2}=2.28$~deg$^{2}$. Our limit of $\rho < 3.9 \times 10^{-3}$ deg$^{-2}$ covers an area of $<2$\,deg$^{2}$ in the most sensitive part of the beam, which is also the most appropriate place to search for transients. As we imaged a larger field of view than the main beam we place limits on these sky areas as well, albeit it with decreased sensitivity. There are however caveats to searching in these far out regions, for example the beam is not well characterised and the ability to detect off-axis transients is not well investigated. 

Furthermore within the main beam region we did not perform deconvolution and so certain areas will be contaminated by bright sources and side lobes and not suitable for transient searches. Our sky area calculations should therefore be considered as conservative upper limits. We note that in calculating the actual noise properties of the images we do account for regions of increased noise due to sources, side lobes and imaging artefacts.  

\begin{table*}
\centering
\caption{The sensitivity calculated over seven discrete sky area bins normalised over all epochs. The corresponding upper limits on the snapshot rates of transients at those areas and sensitivities are given in the final column.}
\begin{tabular}{|c|c|c|c|}
\hline
Cumulative Area  & Sensitivity & 8$\sigma$ Sensitivity & $\rho$ \\
 (deg$^{2}$) &  (Jy beam$^{-1}$) & (Jy beam$^{-1}$) & (deg$^{-2}$)   \\
\hline
2.0 & 0.025 & 0.2  & 3.9$\times 10^{-3}$ \\
4.0 &0.037 & 0.3 & 1.9$\times 10^{-3}$ \\
6.0 & 0.053 & 0.4 & 1.3$\times 10^{-3}$\\
8.0 & 0.08  &  0.6 & 9.7$\times 10^{-4}$\\
10.0 & 0.14 & 1.1 & 7.7$\times 10^{-4}$ \\
12.0 & 0.31 & 2.5 & 6.4$\times 10^{-4}$\\
13.3 & 1.0 &  8.0 &  5.8$\times 10^{-4}$ \\
\hline
\label{area_sens_table}
\end{tabular}
\end{table*}

These upper limits are time independent, i.e., they do not factor in a characteristic timescale of transient phenomena. From the Pipeline~1 results, we can place a 95\% confidence upper limit (Feldman \& Cousins 1999)\nocite{fc99} on the rate of transients with a $\sim$2-minute time-scale of $7\times10^{-6}$/s/deg$^2$. 

The most comparable work to this with regards to the choice of observing frequency is Bannister et al. (2011). In that study, 22 years of archival SUMSS images at 843 MHz were searched for transients on timescales of 12 hours to years above a flux density threshold of 14\,mJy. Two transient sources were detected in their survey yielding a snapshot surface density of  $\rho=1.5 \times 10^{-3}$~deg$^{-2}$.  The surface density of transients found by Bannister et al. (2011) is comparable to the limits placed via this work albeit at a lower detection threshold. The typical cadence of the two surveys is however very different (minutes versus years). Bannister et al. (2011) reported that the detected transients were extra-galactic in origin and had long timescales $\sim$years, that plus the lower detection threshold of their survey means that it would be impossible for them to be detectable with our observations. However, it is noteworthy that we can achieve a very competitive limit on the surface density of events with such a small amount of observing time and with a commissioning array with 1/6th of the final collecting area and 1/30th of the field-of-view of the final ASKAP telescope.

\subsection{Implications for pulsar discovery}

Traditional pulsar surveys are based on recording the signal from the telescope averaged over a relatively short period ($\sim 100\mu$s) with sufficient frequency resolution.  Off-line processing is carried out to de-disperse the time series with a large number of trial dispersion measures.  For each dispersion measure, a Fourier transform is taken to search for a periodic signal.  This method is affected by radio-frequency interference and telescope gain variations.  The sensitivity of the survey to a particular pulsar significantly degrades if the emission is intermittent or if the pulsar has a binary companion.  These traditional searches are also generally only sensitive to pulsars with periods $>$1\,ms and $<$10\,s.  As all the pipelines tested in this paper managed to clearly identify PSR~J1107$-$5907, it is possible that intermittent pulsars could also be discovered in future ASKAP transient searches and we briefly discuss this  prospect here.

We do not know the properties of the population of intermittent pulsars in detail and therefore it is not possible to provide a detailed study of what new pulsars a given survey may find. We know that pulsars can be detected from a single pulse (the RRATS), they can be ``on" for minutes and then ``off" on time scales of hours (such as PSR~J1717$-$4054; Kerr et al. 2014), days (such as PSR B1931+24 which has a 30\,d time scale; Kramer et al. 2006)\nocite{kramer2006} or that they can be ``on" for years and then ``off" for years (e.g., PSR J1832$+$0029; Lorimer et al. 2012)\nocite{lorimer2012}.    

We have simulated pulsars switching between ``on" and ``off" states on various time scales and then determined whether different survey strategies could detect them.  For traditional pulsar surveys we simply require that the signal, when averaged over the entire observation  is above the limiting flux density sensitivity for the survey  ($\sim$30 minutes and 0.2\,mJy for the Parkes multibeam pulsar survey, Manchester et al. 2001\nocite{manchester11}).  We also consider the detectability of a pulsar as a continuum point-source in an ASKAP continuum survey.  We assume a survey similar to that proposed for the Evolutionary Map of the Universe (EMU) survey which will allow a point source detection limit of $\sim 50 \mu$Jy (Norris et al. 2011)\nocite{norris2011}.  Finding a pulsar in such a survey will involve being able to distinguish a pulsar candidate from all other point-sources.  

We also simulate two transient searches.  First, we have a simulated survey similar to the actual survey carried out in this paper;  we assume that the EMU 12\,hour observations are divided into 5 minute snap-shot images and searched for transient emission.  Second, we consider a transient search in which a specific area of sky is only imaged for 5 minutes, but that observation is repeated once per week over a period of five years (labelled here as a VAST-style survey).

The detectability of a pulsar in these surveys depends upon how long the pulsar remains in its ``on" and ``off" states.  If the pulsar switches on a very short time scale (i.e.,  less than a minute) then all surveys will simply identify the pulsar as a non-transient object and the detectability simply depends upon the limiting flux density of the survey.  Such pulsars will not be found in the transient searches.   However, for pulsars similar to J1107$-$5907, that are on for minutes to hours and then off for a few hours, the chance of detection in the traditional observations decreases as the ``on" time-scale decreases.   

\begin{figure}
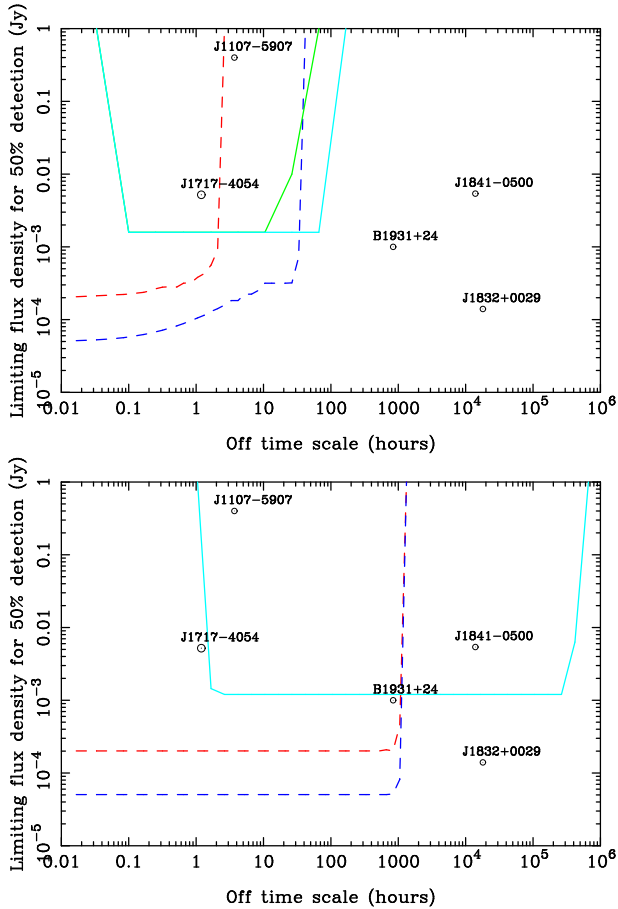

\includegraphics[width=6cm,angle=-90]{result_18m.ps}
\includegraphics[width=6cm,angle=-90]{result_30d.ps}
\caption{Sensitivity of various surveys to intermittent pulsars with different ``off" time scales.  The top panel is for a pulsar that is ``on" for 18\,minutes. The bottom panel is for a pulsar that is ``on" for 30\,days.  The dashed red line indices a  survey similar to the Parkes Multibeam Pulsar Survey. The dashed blue line an EMU-style continuum survey. The solid green line represents a transient survey based on 5 minute snapshots of the 12\,hour EMU observations.  The blue line represents a VAST-style survey in which 5 minute observations are repeated each week for 5 years).   The pulsars indicated have the correct ``off" time scales, but all have different ``on" time scales and are therefore simply included to give the currently known range of switching time scales.}\label{fg:sens1}
\end{figure}

We created a simulated data set in which a pulsar with a given flux density switched ``on" for a certain time and then ``off" for another time.  We then considered 1000 trial searches with specified survey parameters and, in each trial, determined whether the pulsar was detectable or not.  We then increased the pulsar's flux density until it could be detected in at least 50\% of the trials.  This gave us the limiting flux density for 50\% detection for a given ``on" and ``off" time scale.  These results are plotted in Figure~\ref{fg:sens1}.  The top panel is for a pulsar with an ``on" time-scale of 18 minutes (the typical time-scale for PSR~J1717$-$4054).  The bottom panel is for a pulsar with an ``on" time-scale of 30 days (much larger than the observing time for any simulated observation).  In this Figure, the red dashed line indicates the Parkes multibeam pulsar survey, the blue dashed line the EMU ASKAP survey, the green solid line is the transient search based on a single 12\,hour observation and the blue solid line is for repeated, short observations of the same sky position.

The top panel shows that the transient searches cannot detect pulsars with very short ($<$5 minute) ``off" time scales as they always switch ``on" during each 5\,minute snap-shot image and therefore are not observed as transient sources.  These pulsars are detected with the traditional pulsar survey and the EMU survey if their flux densities are greater than the survey sensitivity.    The VAST-style survey provides sensitivity to pulsars with relatively long ``off" time scales, but the survey is only sensitive to relatively bright pulsars.

The bottom panel is more extreme.  For a pulsar that is ``on" for 30\,days, it is unlikely that it switches ``off" during an observation.  Our simulations therefore indicate that such pulsars are unlikely to be detected in a transient search of the 12\,hour EMU observations.  No green, solid line is therefore shown in this panel. The EMU continuum survey and the traditional pulsar survey are limited only by the survey sensitivity until the ``off" time scale reaches $>$30\,days. The VAST-style survey is sensitive to such pulsars with a much greater range of ``off" time scales.

We therefore conclude that repeated, relatively short observations of the same sky region provides the best survey strategy for discovering unknown intermittent pulsars.  We note that the advantages of such a survey with ASKAP would also include 1) observations at a very radio-quiet site (RFI mitigation is currently a major challenge for finding such pulsars) and 2) is at a lower frequency than most Parkes surveys (and therefore pulsars are likely to be brighter).

\section{Conclusion}

We have shown that BETA can produce high-quality images of complex regions of the Galactic plane that are consistent with previous work. The observations allow searches for transient events on time-scales between minutes and $\sim 12$\,hours. These are time-scales that have not previously been studied in detail.  We have shown that pipelines developed for LOFAR and ASKAP can successfully detect PSR~J1107$-$5907, but no other transient sources were detected.

This work has just touched on the possibilities that will arise with the full ASKAP telescope. We have shown that it will be an ideal telescope for discovering intermittent pulsars that spend significant amounts of time in their ``off" state.  This work therefore demonstrates the role of future, wide-area survey telescopes to detect and identify transient sources of radio emission.  It also demonstrates the requirement for a sensitive telescope, like Parkes, that can follow-up on the detected events with wide frequency coverage, high time resolution and the ability to carry out long-term monitoring programs. Together such telescopes will be able to discover and study both expected and unexpected transient sources.

\section{Acknowledgements}

The Australian SKA Pathfinder is part of the Australia Telescope National Facility which is managed by CSIRO. Operation of ASKAP is funded by the Australian Government with support from the National Collaborative Research Infrastructure Strategy. Establishment of the Murchison Radio-astronomy Observatory was funded by the Australian Government and the Government of Western Australia. ASKAP uses advanced supercomputing resources at the Pawsey Supercomputing Centre. We acknowledge the Wajarri Yamatji people as the traditional owners of the Observatory site.  Parts of this research were conducted by the Australian Research Council Centre of Excellence for All-sky Astrophysics (CAASTRO), through project number CE110001020.  The Parkes radio telescope is part of the Australia Telescope, which is funded by the Commonwealth of Australia for operation as a National Facility managed by the Commonwealth Scientific and Industrial Research Organisation (CSIRO). This paper includes archived data obtained through the CSIRO Data Access Portal (http://data.csiro.au). The work was supported by iVEC through the use of advanced computing resources located at The Pawsey Centre. We gratefully acknowledge the LOFAR Transients Key Science Project for providing us with access to the LOFAR Transients Pipeline prior to the public release. 

\bibliography{refs}
\bibliographystyle{mn}

\section{Author Affiliations}
$^{1}$ CSIRO Astronomy and Space Science, PO Box 76, Epping NSW 1710, Australia \\
$^{2}$ Department of Physics and Electronics, Rhodes University, P. O. Box 94, Grahamstown, South Africa \\
$^{3}$ ARC Centre of Excellence for All-sky Astrophysics (CAASTRO) \\
$^{4}$ Sydney Institute for Astronomy, School of Physics, University of Sydney NSW 2006, Australia \\
$^{5}$ International Centre for Radio Astronomy Research (ICRAR), Curtin University, GPO Box U1987, Perth WA 6845, Australia \\
$^{6}$ Radio Astronomy Laboratory, University of California Berkeley, 501 Campbell, Berkeley CA 94720-3411, USA \\
$^{7}$ SKA Organisation, Jodrell Bank Observatory, Lower Withington, Macclesfield Cheshire SK11 9DL, United Kingdom \\
$^{8}$ Inter-University Centre for Astronomy and Astrophysics, Post Bag 4, Ganeshkhind, Pune University Campus, Pune 411 007, India \\
$^{9}$ CSIRO Digital Productivity, PO Box 76, Epping NSW 1710, Australia \\
$^{10}$ Research School of Astronomy and Astrophysics, Australian National University, Mount Stromlo Observatory, Cotter Road, Weston Creek ACT 2611, Australia \\
$^{11}$ International Centre for Radio Astronomy Research (ICRAR), University of Western Australia, 35 Stirling Highway, Crawley WA 6009, Australia \\
$^{12}$ Sonartech ATLAS Pty Ltd,  Unit G01, 16 Giffnock Avenue, Macquarie Park NSW 2113, Australia \\
$^{13}$ School of Physics, University of Melbourne, VIC, 3010, Australia \\
$^{14}$ Leiden Observatory, Leiden University, PO Box 9513, NL-2300 RA  Leiden, The Netherlands \\
$^{15}$ 31 Ellalong Road, North Turramurra NSW  2074, Australia \\
$^{16}$ Anton Pannekoek Institute, University of Amsterdam, Postbus 94249, 1090 GE, Amsterdam, The Netherlands \\
$^{17}$ Netherlands Institute for Radio Astronomy (ASTRON), PO Box 2, 7990 AA Dwingeloo, The Netherlands \\
\end{document}